\documentclass[runningheads]{llncs}
\usepackage[T1]{fontenc}
\usepackage{multirow}
\usepackage{booktabs}
\usepackage{graphicx}
\usepackage{hyperref}
\usepackage{bbm} % cz add
\usepackage{amssymb}
\usepackage{wrapfig} % cz add
\usepackage{array}
\usepackage{bm}
\usepackage{amsmath}
\usepackage{breakcites}
\usepackage[misc]{ifsym}

\begin{document}
\title{DiffRect: Latent Diffusion Label Rectification for Semi-supervised Medical 
Image Segmentation}
\titlerunning{Latent Diffusion Label Rectification for Semi-supervised Medical Image}
\author{Xinyu Liu \and
Wuyang Li \and
Yixuan Yuan$^{\text{\Letter}}$} 
% index{Liu, Xinyu}
% index{Li, Wuyang}
% index{Yuan, Yixuan}
%
\authorrunning{Liu, Li, and Yuan}
\institute{Department of Electronic Engineering, The Chinese University of Hong Kong, Shatin, Hong Kong SAR\\
\email{yxyuan@ee.cuhk.edu.hk}}

\maketitle              % typeset the header of the contribution
\begin{abstract}

Semi-supervised medical image segmentation aims to leverage limited annotated data and rich unlabeled data to perform accurate segmentation. However, existing semi-supervised methods are highly dependent on the quality of self-generated pseudo labels, which are prone to incorrect supervision and confirmation bias. Meanwhile, they are insufficient in capturing the label distributions in latent space and suffer from limited generalization to unlabeled data.
To address these issues, we propose a Latent Diffusion Label Rectification Model (DiffRect) for semi-supervised medical image segmentation. 
DiffRect first utilizes a Label Context Calibration Module (LCC) to calibrate the biased relationship between classes by learning the category-wise correlation in pseudo labels,
then apply Latent Feature Rectification Module (LFR) on the latent space to formulate and align the pseudo label distributions of different levels via latent diffusion. 
It utilizes a denoising network to learn the coarse to fine and fine to precise consecutive distribution transportations. 
We evaluate DiffRect on three public datasets: ACDC, MS-CMRSEG 2019, and Decathlon Prostate. Experimental results demonstrate the effectiveness of DiffRect, e.g. it achieves 82.40\% Dice score on ACDC with only 1\% labeled scan available, outperforms the previous
state-of-the-art by 4.60\% in Dice, and even rivals fully supervised performance. Code is released at \url{https://github.com/CUHK-AIM-Group/DiffRect}.

\end{abstract}
% \vspace{-0.3cm}
\section{Introduction}
Medical image segmentation is crucial for clinical applications but often requires large amounts of pixel-wise or voxel-wise labeled data, which is tedious and time-consuming to obtain \cite{liu2021consolidated, liu2022source, liu2023decoupled, bai2017semi_selftrain,sun2022few}. Such a heavy annotation cost has motivated the community to develop semi-supervised learning methods~\cite{jiao2022ss4mis_survey, ssl4mis2020,li2022hierarchical,yangqiushi2022semi}. 
Existing semi-supervised image segmentation methods can be generally categorized into self-training and consistency regularization. For self-training methods~\cite{bai2017semi_selftrain, vu2019advent, chen2019multi, feng2020semiselftraining, wang2022semiu2pl,li2021consistent, zhang2021self, mendel2023error}, they generate pseudo labels for unlabeled images, then use the pseudo-labeled images in conjunction with labeled images to update the segmentation model iteratively. This paradigm could effectively incorporate
unlabeled data by minimizing their entropy. For consistency regularization methods~\cite{crosspseudosupervision, sohn2020fixmatch, luo2022semicnntransformer, verma2022interpolationconsistencytraining, wu2022mutual, AEL, luo2022semipyramid, yang2023revisiting, wang2023mcf}, they are designed based on the assumption that perturbations should not change the predictions of the model, and have achieved more promising performance recently. Perturbations are applied on the input or the network level, and the models are enforced to achieve an invariance of predictions.

Despite the progress, the semi-supervised medical image segmentation remains challenging due to the following factors.
(1) \textbf{Reliance Risk}: Existing methods typically rely on self-generated pseudo labels to optimize the model~\cite{verma2022interpolationconsistencytraining, yang2023revisiting,li2022knowledge,wu2022mutual}, which is ill-posed since errors in pseudo labels are preserved during iterative optimization. The overfitting to incorrect supervision could lead to severe confirmation bias~\cite{li2019dividemix} and considerable performance degradation. Besides, they do not fully utilize the category-wise correlation in the pseudo labels, and the label quality is sensitive to the perturbation design and network structure.
(2) \textbf{Distribution Misalignment}: Most methods only apply consistency regularization and auxiliary supervision at the output mask level to encourage the model to produce consistent mask predictions between different perturbations~\cite{sohn2020fixmatch, crosspseudosupervision}. However, these approaches are insufficient in capturing the semantics in the latent space and tend to overlook the underlying label distributions,
resulting in limited generalization to unlabeled data.

To address the reliance risk issue,  we first propose a \textbf{L}abel \textbf{C}ontext \textbf{C}alibration Module (LCC). Different from methods that directly use the self-generated pseudo labels, LCC 
calibrates the biased semantic context, \textit{i.e.}, the relationships between different semantic categories, and reduce the errors in the pseudo labels. It starts with a semantic coloring scheme that encodes the one-hot pseudo labels and ground truth masks into the visual space, and subsequently feeds them into a semantic context embedding block to adjust the features of the pseudo labels in the latent space. Notably, LCC introduces explicit calibration guidance by encoding the dice score between the pseudo labels and the ground truth, thereby providing more reliable calibration directions for model optimization.%

To tackle the distribution misalignment problem, some previous works have proposed to model data distributions with VAE~\cite{VAEssl_2019} or GAN~\cite{zhang2021generator}. However, their adversarial training scheme could suffer from mode collapse and conflict between generation and segmentation tasks, resulting in suboptimal performance. Different from them, the denoising diffusion probabilistic model (DDPM) is a new class of generative models 
trained using variational inference~\cite{ho2020denoising, nichol2021improvedddpm, li2024endora,li2024u}, which alleviates the above problem by formulating the complex data distribution with probabilistic models. Therefore, 
we design a \textbf{L}atent \textbf{F}eature \textbf{R}ectification Module (LFR), 
which models the consecutive refinement between different latent distributions 
with a generative latent DDPM~\cite{latentdiffusion}. 
LFR leverages the power of DDPM to learn the latent structure of the semantic labels. 
Specifically, it first applies Gaussian noise on fine-grained label features with a diffusion schedule,
then uses the coarse-grained label features as conditions to recover the clean feature. With the denoising process, the consecutive transportations of coarse to fine and fine to precise distributions of the pseudo labels are formulated and aligned, and the pseudo labels are progressively rectified for better supervision. 
Based on LCC and LFR, we construct a semi-supervised medical image segmentation framework named Latent \textbf{Diff}usion Label \textbf{Rect}ification Model (\textit{DiffRect}). Extensive experimental results show that our method outperforms prior methods by significant margins.

% \vspace{-2pt}
\section{Methodology}
\subsection{Preliminary: Conditional DDPM}
DDPM is a class of latent variable generative model that learns a data distribution by denoising noisy images~\cite{ho2020denoising}. The forward process diffuses the data samples with pre-defined noise schedules. Concretely, given a clean data $z^0$, sampling of $z^t$ is expressed in a closed form: 
\begin{equation}
    q(z^t\|z^0) = \mathcal{N}(z^t; \sqrt{\overline{\alpha}_t}z^0, (1-\overline{\alpha}_t)\mathbf{I}),
\end{equation}
where $\overline{\alpha}_t$ is the noise schedule variable~\cite{nichol2021improvedddpm, ho2020denoising}. During the reverse process, we are given an optional condition $\rho$~\cite{choi2021ilvr}, and each step is expressed as a Gaussian transition with learned mean $\boldsymbol{\mu}_\epsilon$ and variance $\sigma_\epsilon$ from the denoising model $\epsilon$:
\begin{equation}
\scalebox{0.93}{
$
    p\left(z^{t-1} \mid z^t, \rho\right):=\mathcal{N}\left(z^{t-1} ; \boldsymbol{\mu}_\epsilon\left(z^t, t, \rho\right), \sigma_\epsilon\left(z^t, t, \rho\right) \mathbf{I}\right).
$
}
\end{equation}
By decomposing the above equation, we have:
\begin{equation}
    z^{t-1} \leftarrow \frac{1}{\sqrt{\alpha_t}}(z^{t} - \frac{1-\alpha_t}{\sqrt{1-\overline{\alpha}_t}}\epsilon(z^{t}, t, \rho)) + \sigma_{\epsilon}\eta,
\end{equation}
where $\eta\sim\mathcal{N}(\mathbf{0}, \mathbf{I})$ is a sampled noise that ensures each step is stochastic. In this work, we extend the conditional DDPM to the latent space of pseudo labels, and model the distribution transportations for label rectification.

\subsection{Label Context Calibration Module (LCC)}

\begin{figure*}[t]
    \centering
    \includegraphics[width=0.93\textwidth]{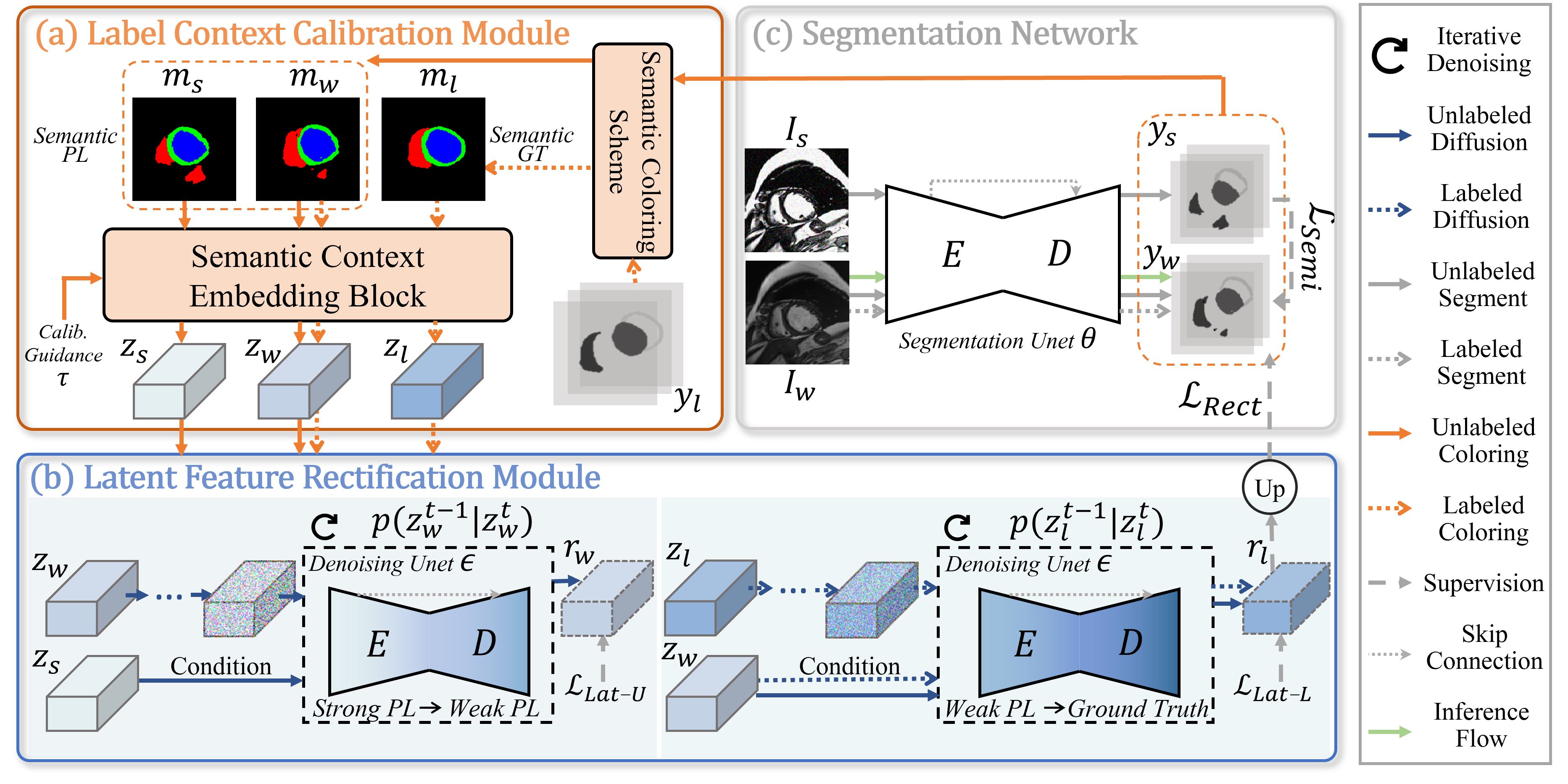}
    % \vspace{-5pt}
    \caption{Overall framework of DiffRect. (a) Label Context Calibration Module (LCC). (b) Latent Feature Rectification Module (LFR). (c) Segmentation Network.}
    % \vspace{-5pt}
    \label{fig:main}
\end{figure*}

Existing semi-supervised training schemes that rely extensively on self-generated pseudo labels are often ill-posed, where errors in low-quality pseudo labels accumulate and degrade performance. To address this issue, we introduce LCC that effectively captures and calibrates the semantic context within the visual space, thereby mitigating the impact of noisy labels. As in Fig. \ref{fig:main}(a), given the one-hot pseudo labels $y_s, y_w \in \mathbb{R}^{H \times W \times C}$ with height $H$ and width $W$ from the segmentation network, we encode them to semantic pseudo labels $m_s$ and $m_w$ with dimensions of $\mathbb{R}^{H \times W \times 3}$, using a proposed \textit{semantic coloring scheme (SCS)}. 

Concretely, for a dataset that contains $C$ different classes, we build a color set $M_C$ that is composed of $C$ RGB colors, and each color is represented by a tuple of three values within the range $[0, 255]$. We maximize the color difference between each encoded category to avoid semantic confusion. Therefore, it can be represented by a functional mapping $f:C \to M_C$, which is defined as:
\begin{equation}
\scalebox{0.9}{
$
    m_{(h,w)} = f(y_{(h,w)}), \quad {\forall}h\in [1, 2, ..., H], w \in [1, 2, ..., W],
    $
    }
\end{equation}
where $m$ is the semantic pseudo label in the visual space, and $m_{(h,w)}$ represents the mapped RGB color of the pixel at location $(h,w)$ in $m$. The $y_{(h,w)}$ represents the class of the corresponding pixel in one-hot mask $y$. The semantic coloring scheme can effectively incorporate color information into the segmentation task, which enables the model to exploit additional cues with the rich semantics from colors, and improves the discrimination ability~\cite{wang2023images, chen2023generativess} as well as the interpretability of the model.

To perform context calibration with the semantic labels, we design a semantic context embedding block $\mathbf{B}_{sem}$, which embeds the pseudo labels to the latent features $z_s, z_w, z_l$ with the dimensions of $\mathbb{R}^{\frac{H}{16} \times \frac{W}{16} \times 256}$. Specially, additional \textit{calibration guidance (CG)} $\tau^u$ for unlabeled data and $\tau^l$ for labeled data are also encoded into the block using the sinusoidal embeddings~\cite{ho2020denoising, vaswani2017attention},

\begin{equation}
\begin{aligned}
    \{z_s, z_w\} &= \mathbf{B}_{\text{sem}}(m_s, m_w \| \tau^u) \quad \text{for unlabeled data}, \\
    \{z_w, z_l\} &= \mathbf{B}_{\text{sem}}(m_w, m_l \| \tau^l) \quad \text{for labeled data}, 
\end{aligned}
\label{eq:sememb}
\end{equation}
where the $\tau^u$ and $\tau^l$ values for unlabeled and labeled data are computed using the dice coefficient between the one-hot segmentation masks of different qualities,
which is denoted as follows:
\begin{equation}
    \tau^u = \text{Dice}(y_s, y_w), \quad \tau^l = \text{Dice}(y_w, y_l).
\label{eq:tau}
\end{equation}
By using the dice coefficient as the calibration guidance factor, the model can simultaneously measure the quality of pseudo labels and integrate this information into the learning process. It enables the model to better capture the semantic context and refine the pseudo labels for both unlabeled and labeled data. 

\subsection{Latent Feature Rectification Module (LFR)}

To address the distribution misalignment issue between the pseudo labels with different levels of quality, 
we propose a Latent Feature Rectification Module (LFR), which is illustrated in Fig. \ref{fig:main}(b).

Concretely, LFR applies a latent diffusion process to model the transportation of label quality distributions. For each unlabeled data $I_u$, the strongly and weakly semantic context embedding $z_s$ and $z_w$ are first obtained with LCC. We then construct a diffusion process from $z_w$ to the diffused noisy feature $z^T_w$ with $T$ timestamps as follows:
\begin{equation}
\begin{aligned}
    z^T_w &= \sqrt{\alpha_T}z^{T-1}_w + \sqrt{1 - \alpha_T}\eta^{T-1} \\
    &=\cdot \cdot \cdot = \sqrt{\overline{\alpha}_T}z_w + \sqrt{1 - \overline{\alpha}_T}\eta,
\end{aligned}
\end{equation}
where $\alpha_T$ and $\overline{\alpha}_T$ are the schedule variables in the diffusion forward process, (\textit{e.g.}, cosine~\cite{nichol2021improvedddpm}), and $\overline{\alpha}_T = \prod^T_{i=1} \alpha_i$. The $\eta^t$ is the corresponding noise sampled from Gaussian distribution at the $t$-th step.
Then, we train a denoising U-Net $\epsilon$ to learn to reverse this process. Since the individual reverse diffusion process is unconditioned, we add $z_s$ as the conditional input and also feed it into the denoising model. Therefore, the model is encouraged to learn the distribution transportation from \textit{coarse-grained masks $p(z_s)$ (strong pseudo labels)} to the latent distributions of \textit{fine-grained masks $p(z_w)$ (weak pseudo labels)}, where we denote it as a \textit{strong-to-weak transportation (S2W)}. The reverse diffusion is formulated as the following Markov chain:
% \vspace{-5pt}
\begin{equation}
\scalebox{0.85}{
$
\begin{aligned}
& p_\epsilon\left(z_w^{0: T}\right):=p\left(z_w^T\right) \prod_{t=1}^T p_\epsilon\left(z_w^{t-1} \mid z_w^t, z_s\right), \quad z_w^T \sim \mathcal{N}(\mathbf{0}, \mathbf{I}) \\
& p_\epsilon\left(z_w^{t-1} \mid z_w^t, z_s\right):=\mathcal{N}\left(z_w^{t-1} ; \boldsymbol{\mu}_\epsilon\left(z_w^t, t, z_s\right), \sigma_\epsilon\left(z_w^t, t, z_s\right) \mathbf{I}\right),
\end{aligned}
$
}
\end{equation}
where $\boldsymbol{\mu}$ and $\sigma$ are the predicted data mean and variance from the denoising U-Net model. For the training with unlabeled input, the \textit{latent loss} for optimization can be expressed as follows:
\begin{equation}
    \mathcal{L_{\text{Lat-U}}}=E_{z_w, t}\left[\left\|z_w-r_w\right\|_2\right],
    \label{lat_u_loss}
\end{equation}
where $r_w = \epsilon\left(z_w^T, z_s, t\right)$,
% \begin{equation}
    % \end{equation}
which is the reconstructed version of the weakly semantic context embedding $z_w$. The objective minimizes the $\ell_2$ distance between the clean and denoised feature and encourages the model to learn the distribution transportation from a coarse pseudo label to a fine pseudo label.

Similarly, we can obtain the weak semantic context embedding of labeled data $z_w$ and the ground truth $z_l$. We then learn the reverse process that recovers $z_l$ based on the $T$-timestamp diffused noisy feature $z_l^T$, with the $z_w$ as condition:
% \vspace{-8pt}
\begin{equation}
\scalebox{0.85}{
$
\begin{aligned}
& p_\epsilon\left(z_l^{0: T}\right):=p\left(z_l^T\right) \prod_{t=1}^T p_\epsilon\left(z_l^{t-1} \mid z_l^t, z_w\right), \quad z_l^T \sim \mathcal{N}(\mathbf{0}, \mathbf{I}) \\
& p_\epsilon\left(z_l^{t-1} \mid z_l^t, z_w\right):=\mathcal{N}\left(z_l^{t-1} ; \boldsymbol{\mu}_\epsilon\left(z_l^t, t, z_w\right), \sigma_\epsilon\left(z_l^t, t, z_w\right) \mathbf{I}\right),
\end{aligned}
$
}
\end{equation}
and the training objective for the reconstructed feature $r_l = \epsilon\left(z_l^T, z_w, t\right)$ is:
\begin{equation}
    \mathcal{L_{\text{Lat-L}}} =E_{z_l, t}\left[\left\|z_l-r_l\right\|_2\right].
    \label{lat_l_loss}
\end{equation}
With the above latent diffusion process, the continual distribution transportations from \textit{fine-grained mask distributions $p(z_w)$ (weak pseudo labels)} to \textit{precise mask distributions $p(z_l)$ (ground truth)} are also formulated in the latent space, which is denoted as the \textit{weak-to-groud truth transportation (W2G)}. The denoising U-Net is hence capable to achieve latent feature rectification.

Afterwards, the weak pseudo labels of unlabeled data are fed into the denoising U-Net for obtaining the rectified features with progressive denoising. Specifically, we randomly sample a Gaussian noise $r_l^{T} \sim \mathcal{N}(\mathbf{0}, \mathbf{I})$ as the input of the denoising U-Net, which simulates the $T$-timestamp noisy feature of the rectified pseudo label $y_r$. The rectified feature $r_l$ is generated via a progressive reverse diffusion process, with the weak pseudo label features $z_w$ as condition. Mathematically, a single denoising from step $t$ to $t-1$ is formulated as:
\begin{equation}
    r_l^{t-1} \leftarrow \frac{1}{\sqrt{\alpha_t}}(r_l^{t} - \frac{1-\alpha_t}{\sqrt{1-\overline{\alpha}_t}}\epsilon(r_l^{t}, t, z_w)) + \sigma_{\epsilon}\eta, 
    \label{lfr_each_step_eq}
\end{equation}
where $\eta\sim\mathcal{N}(\mathbf{0}, \mathbf{I})$ which ensures each step is stochastic as in DDPM~\cite{ho2020denoising}. The rectified label is obtained with an upsampling of the feature $r_l$ to the input resolution $y_r = Upsample(r_l)$, which is utilized as a better and more precise supervision signal for the segmentation model.

\subsection{Loss Function}

The training of the DiffRect frameworks includes two parts: (1) the optimization of segmentation U-Net $\theta$ (with Seg Loss) and (2) the joint optimization of the rectification components $\mathbf{B}_{sem}$ and $\epsilon$ (with Diff Loss). The overall loss is:
\begin{equation}
\scalebox{0.9}{
$
\mathcal{L_{\text{DiffRect}}} = \underbrace{{\mathcal{L^{\text{Seg}}_{\text{Semi}}} + \mathcal{L_{\text{Rect}}}}}_{\text{Seg Loss}} + \underbrace{{\mathcal{L^{\text{Lat}}_{\text{Semi}}} + \lambda_1 \mathcal{L_{\text{Lat-U}}} + \lambda_2 \mathcal{L_{\text{Lat-L}}}}}_{\text{Diff Loss}},
$
}
\label{all_losses_eq}
\end{equation}
where $\mathcal{L^{\text{Seg}}_{\text{Semi}}}$ and $\mathcal{L^{\text{Lat}}_{\text{Semi}}}$ are the semi-supervised losses for segmentation as in~\cite{sohn2020fixmatch}. The $\lambda_1$ and $\lambda_2$ are trade-off factors to balance the contribution of each term. $\mathcal{L_{\text{Rect}}}$ is the rectified supervision loss between $y_w$ and the rectified pseudo label $y_r$, where the summation of cross-entropy and Dice score are used:
\begin{equation}
    \mathcal{L_{\text{Rect}}} = \text{CE}(y_w, y_r) + \text{Dice}(y_w, y_r).
\end{equation}
During inference, the input is directly fed into segmentation network in Fig. \ref{fig:main}(c) to produce the segmentation result, thus no extra inference cost is required.

\begin{table*}[t]
  \centering
  \caption{Segmentation results on the ACDC validation and test sets.}
   % $\uparrow$ denotes the higher the better and vice versa. All experiments are performed in identical settings for fair comparisons. Bold refers to the best performance.
    \setlength{\tabcolsep}{1.2mm}
    \scalebox{0.84}{
    \begin{tabular}{c|c|cccc|cccc}
    \hline
    \multirow{2}[1]{*}{Method} & Labeled & \multicolumn{4}{c|}{ACDC Validation Set} & \multicolumn{4}{c}{ACDC Test Set} \\
\cline{3-10}          & Ratio   & Dice$\uparrow$ & Jac$\uparrow$ & HD95$\downarrow$ & ASD$\downarrow$& Dice$\uparrow$ & Jac$\uparrow$ & HD95$\downarrow$ & ASD$\downarrow$ \\
    \hline
    UAMT~\cite{yu2019uncertaintyawaremeanteacher} & \multirow{7}[1]{*}{1\%}  & 42.28 & 32.21 & 40.74 & 18.58 & 43.86 & 33.36 & 38.60 & 18.33 \\
    FixMatch~\cite{sohn2020fixmatch} &   & 69.67  & 58.34 & 37.92& 14.41 & 60.80  & 49.14 & 36.81& 14.75 \\
    CPS~\cite{crosspseudosupervision} &  & 56.70 & 44.31& 24.97 & 10.48& 52.28  & 41.68 & 20.38& 7.35 \\
    ICT~\cite{verma2022interpolationconsistencytraining}  &   & 43.03 & 30.58 & 34.92 & 15.23 & 42.91 & 32.81 & 25.42 & 10.80 \\
    MCNetV2~\cite{wu2022mutual} &  & 57.49 & 43.29 & 31.31 & 10.97 & 49.92  & 39.16& 24.64& 8.47 \\
    INCL~\cite{zhu2023inherentINCL} &  & 77.80 & 66.13 & 11.69 & 3.22 & 67.01 & 56.22 & 13.43 & 3.35 \\
    \textbf{DiffRect (Ours)} &   & \textbf{82.40} & \textbf{71.96}& \textbf{10.04}  & \textbf{2.90} & \textbf{71.85} & \textbf{61.53} & \textbf{5.79} & \textbf{2.12} \\
    \hline % 3 LABELED
    UAMT~\cite{yu2019uncertaintyawaremeanteacher} &  \multirow{7}[1]{*}{5\%}  & 72.71 & 60.89 & 21.48 & 7.15 & 69.93  & 58.45& 17.01 & 5.25 \\
    FixMatch~\cite{sohn2020fixmatch}&   & 83.12 & 73.59 & 9.86 & 2.61 & 74.68  & 64.12 & 11.18& 2.93 \\
    CPS~\cite{crosspseudosupervision} &  & 75.24 & 64.67 & 10.93 & 2.98 & 74.67  & 63.51 & 9.37& 2.55 \\
    ICT~\cite{verma2022interpolationconsistencytraining}  &   & 74.20  & 62.90 & 17.01& 4.32 & 73.10  & 60.69& 11.92 & 3.70 \\
    MCNetV2~\cite{wu2022mutual} &  & 78.96 & 68.15 & 12.13 & 3.91 & 75.86 & 65.20 & 9.85 & 2.88 \\
    INCL~\cite{zhu2023inherentINCL} &  & 85.43 & 75.76 & 6.37 & 1.37 & 80.64 & 70.78 & \textbf{5.29} & \textbf{1.42}\\
    \textbf{DiffRect (Ours)} &   & \textbf{86.95}  & \textbf{78.08} & \textbf{4.07}& \textbf{1.23} & \textbf{82.46}  & \textbf{71.76} & {7.18}& {1.94} \\
    \hline
    UAMT~\cite{yu2019uncertaintyawaremeanteacher} & \multirow{7}[1]{*}{10\%}  & 85.14 & 75.90& 6.25   & 1.80& 86.23   & 76.72& 9.40 & 2.56\\
    FixMatch~\cite{sohn2020fixmatch}&   & 88.31   & 79.97 & 7.35& 1.79 & 87.96   & 79.37 & 5.43& 1.59 \\
    CPS~\cite{crosspseudosupervision} &  & 84.63  & 75.20& 7.57 & 2.27 & 85.61 & 75.76 & 9.29& 3.00 \\
    ICT~\cite{verma2022interpolationconsistencytraining}  &   & 85.15 & 76.05 & 4.27  & 1.46 & 86.77  & 77.43 & 8.01 & 2.16 \\
    MCNetV2~\cite{wu2022mutual} &  & 85.97 & 77.21 & 7.55 & 2.11 & 88.75 & 80.28 & 6.16 & 1.64 \\
    INCL~\cite{zhu2023inherentINCL} &  & 88.28 & 80.09 & 1.67 & 0.49 & 88.68 & 80.27 & 4.34 & 1.13\\
    \textbf{DiffRect (Ours)} &   & \textbf{90.18}  & \textbf{82.72} & \textbf{1.38}& \textbf{0.48} & \textbf{89.27}  & \textbf{81.13} & \textbf{3.85}& \textbf{1.00} \\
    \hline
    Supervised~\cite{unet}& 100\%  & 91.48  & 84.87 & 1.12 & {0.34}  & 91.65   & 84.95 & 1.14& 0.50 \\
    \hline
    % \bottomrule
    \end{tabular}}%
  \label{tab:addlabel}%
\label{table_acdc}
% \vspace{-6pt}
\end{table*}%

\begin{table}[t]
\begin{minipage}{0.48\linewidth}
\centering
\caption{Segmentation results on MS-CMRSEG 2019 with 20\% data labeled.}
\resizebox{0.99\textwidth}{!}{\begin{tabular}{c|cccc}
% \small
    \hline
 Method       & Dice $\uparrow$ & Jac$\uparrow$ & HD95$\downarrow$ & ASD$\downarrow$ \\
    % \midrule
    \hline
    UAMT~\cite{yu2019uncertaintyawaremeanteacher}   & 
    84.27	&73.69	&12.15	&4.18
    \\
    FixMatch~\cite{sohn2020fixmatch}& 84.31 &	73.57 & 	17.79 &	4.81
 \\
    CPS~\cite{crosspseudosupervision}& 83.66 &73.03 &15.01&	4.30
\\
    ICT~\cite{verma2022interpolationconsistencytraining}  & 83.66	&73.06&	17.24&	4.85
  \\
    MCNetV2~\cite{wu2022mutual}&  83.93	&73.45	&13.10	&3.39 \\
    INCL~\cite{zhu2023inherentINCL} & 84.33 &	73.92	& 9.95 &	2.61\\
    \textbf{DiffRect} &  \textbf{86.78} & \textbf{77.13} & \textbf{6.39} & \textbf{1.85} \\
    \hline
    Supervised~\cite{unet}&  88.19	&79.28	&4.21	&1.32 \\
    % \\
    \hline
    % \bottomrule
    \end{tabular}}\label{tab:mscmrseg19}
\end{minipage}
\hspace{1.2em}
\begin{minipage}{0.48\linewidth}
\centering
\caption{Segmentation results on Decathlon Prostate with 10\% data labeled.}
\resizebox{0.99\textwidth}{!}{\begin{tabular}{c|cccc}
    \hline
Method  & Dice$\uparrow$ & Jac$\uparrow$ & HD95$\downarrow$ & ASD$\downarrow$ \\
    \hline
    UAMT~\cite{yu2019uncertaintyawaremeanteacher} & 40.91&	29.13&	28.32&	10.45				
    \\
    FixMatch~\cite{sohn2020fixmatch}& 54.70&	41.07&	16.82&	5.24
 \\
    CPS~\cite{crosspseudosupervision}& 43.51&	31.18	&26.93	&8.31
\\
    ICT~\cite{verma2022interpolationconsistencytraining}  & 39.91&	28.95	&24.73&	7.59
 \\
    MCNetV2~\cite{wu2022mutual}& 40.58	&28.77	&21.29	&7.11
\\
    INCL~\cite{zhu2023inherentINCL} & 55.67&41.91	&31.09	&15.78				\\
    \textbf{DiffRect} & \textbf{62.23}&	\textbf{48.64}&	\textbf{10.36}&	\textbf{3.41}
\\
    \hline
    Supervised~\cite{unet}  & 73.81&	61.25&	7.28&	1.94
    \\
    \hline
    % \bottomrule
    \end{tabular}}\label{tab:decathlon}
\end{minipage}
% \vspace{-6pt}
\end{table}

\section{Experiments}
\label{sec:experiments}

\subsection{Experimental Setup}
\label{ssec:datasets}

We examine all methods with identical settings for fair comparison, and trained on a NVIDIA 4090 GPU for 30k iterations. For the $\mathbf{B}_{sem}$ which downsamples the input to $\frac{H}{16} \times \frac{W}{16}$, we use two $3 \times 3$ convolution layers followed by BN and LeakyReLU before the 2$\times$ downsample in each stage, and repeat for four stages. The Denoising U-Net $\epsilon$ down and upsamples the input by 4$\times$, which also uses two $3 \times 3$ convolution layers per stage. The multi-scale image feature is embedded into the model via concatenation as in~\cite{xing2023diff_u_net}. 
For the weak perturbation, we apply random flipping and rotation. For the strong perturbation, we apply random Gaussian blur and additional random image adjustments, including contrast, sharpness, and brightness enhancement. 
For ACDC, we test the 1\%, 5\%, and 10\% labeling regimes following \cite{ssl4mis2020}. For MS-CMRSEG 2019, 20\% labeling regime is tested, while 10\% labeled data is used in Decathlon Prostate. 

\subsection{Comparison with State-of-the-art Methods}

We validate the effectiveness of the proposed approach on the ACDC dataset~\cite{ACDC} in Tab. \ref{table_acdc}.
Our method shows superior results under all labeling regimes. 
Compared with MCNetV2~\cite{wu2022mutual}, our method possesses superior capability with increments of 24.91\%, 7.99\%, 4.21\% in Dice, 28.67\%, 9.93\%, 5.51\% in Jaccard on the validation set with 1\%, 5\%, and 10\% scans available. 
DiffRect displays better segmentation performance even when the labeled samples are extremely scarce (\textit{e.g.} 82.40\% Dice with 1\% scans available), suggesting it can model the transportation of the pseudo label distributions precisely and produce refined masks. Results in MS-CMRSEG 2019 are shown in Tab. \ref{tab:mscmrseg19}. DiffRect shows consistent performance gain on all metrics, with 86.78\% in Dice, 77.13\% in Jaccard, 6.39mm in HD95, and 1.85mm in ASD, outperforming the state-of-the-art method INCL~\cite{zhu2023inherentINCL} by 2.45\% Dice, 3.21\% Jaccard, 3.56mm HD95, and 0.76mm in ASD, respectively. On Decathlon Prostate in Tab. \ref{tab:decathlon}, DiffRect remains showing compelling results, demonstrating its capability in various modalities.

\begin{table}[t]
\begin{minipage}{0.48\linewidth}
\centering
\caption{Ablation study of the proposed modules.}
\resizebox{0.99\textwidth}{!}{\begin{tabular}{l|c|cccc}
    \hline
    % \toprule
Method   & w/o & Dice$\uparrow$ & Jac$\uparrow$ & HD95$\downarrow$ & ASD$\downarrow$ \\
\hline
% \midrule
Baseline & -   & 69.67     & 58.34        & 37.92     & 14.41    \\
\hline
+LCC     & SCS & 73.83     & 61.83        & 29.49     & 11.71    \\
         & CG  & 76.12     & 64.69        & 26.24     & 8.31     \\
         & -   & 78.28     & 66.97        & 20.46     & 5.60     \\
        \hline
+LCC     & S2W &  79.97         &  69.31            & 14.07          & 4.91         \\
\& LFR     & W2G &  78.57         &  66.38            & 21.07          & 5.91         \\
         & -   & \textbf{82.40}     & \textbf{71.96}        &\textbf{10.04}     & \textbf{2.90}   \\
    % \bottomrule
    \hline
    \end{tabular}}\label{tab:abl_modules}
\end{minipage}
\hspace{1.2em}
\begin{minipage}{0.48\linewidth}
\centering
\caption{Ablation study of different calibration guidance choices in LCC.}
\resizebox{0.99\textwidth}{!}{\begin{tabular}{l|cccc}
    \hline
    % \toprule
Choice & Dice$\uparrow$ & Jac$\uparrow$ & HD95$\downarrow$ & ASD$\downarrow$ \\
% \midrule
\hline
Dice   & \textbf{82.40}     & \textbf{71.96}        & \textbf{10.04}     & 2.90    \\
Jaccard    &  82.37 & 71.82 & 11.33 & {2.87} \\
Fixed     &  80.34 & 69.67 & 14.97 & 4.47 \\
Random    & 80.60 & 69.99 & 13.15 & 3.75 \\ 
Both &  81.67 & 71.45 & 10.28 & \textbf{2.47}   \\
    % \bottomrule
    \hline
    \end{tabular}}\label{table_abl_loss}
\end{minipage}
% \vspace{-10pt}
\end{table}

\subsection{Further Analysis}
\textbf{Ablation study of the proposed modules.} 
We evaluate the effect of individual modules in DiffRect in Tab. \ref{tab:abl_modules}. Adopting LCC achieves 78.28\% Dice and 66.97\% Jaccard, with 8.61\% and 8.63\% gains compared with the Fixmatch baseline~\cite{sohn2020fixmatch}.
Removing the semantic coloring scheme (SGS) shows a large performance drop (73.83\% Dice and 61.83\% Jaccard), showing the importance of exploiting the semantics in the visual domain. No calibration guidance (CG) causes 2.16\% Dice drop due to the impact of noisy calibration directions. 
Adding LFR improves Dice by 4.12\% and 10.42mm in HD95.
Removing the strong to weak transportation (S2W) shows a 2.43\% Dice drop while removing the weak to ground truth (W2G) causes a severe Dice drop to 78.57\%. The results demonstrate the necessity of 
each sub-component.

\textbf{Different Calibration Guidance Choices.} To analyze the effectiveness and the optimal choice of calibration guidance, experiments were conducted to compare the performance of models trained with different calibration guidance in Tab. \ref{table_abl_loss}, including Dice score, Jaccard score, Fixed (using a fixed value 0.5), Random (using a random sampled value within 0$\sim$1), and Both (using the summation of Dice and Jaccard). It is shown that Dice, Jaccard, and Both have similar performance, and outperform the fixed and random strategies, which validates the reliable directions provided for optimization.

\section{Conclusion}
In this paper, we identify the reliance risk and distribution misalignment issues in semi-supervised medical image segmentation, and propose DiffRect, a diffusion-based framework for this task. It comprises two modules: the LCC aims to calibrate the biased relationship between classes in pseudo labels by learning category-wise correlation, and the LFR models the consecutive transportations between coarse to fine and fine to precise distributions of the pseudo labels accurately with latent diffusion. Extensive experiments on three datasets demonstrate that DiffRect outperforms existing methods by remarkable margins. 

\begin{credits}
\subsubsection{\ackname} This work was supported by Hong Kong Research Grants Council (RGC) General Research Fund 14204321.

% \subsubsection{\discintname}
% The authors have no competing interests to declare that are
% relevant to the content of this article.
\end{credits}

\bibliographystyle{splncs04}
\bibliography{ref}




% LNCS Homepage, \url{http://www.springer.com/lncs}. Last accessed 4
% \end{thebibliography}
\end{document}